\documentclass[
 twocolumn,
]{ceurart}

\usepackage{listings}
\begin{document}

\copyrightyear{2023}
\copyrightclause{Copyright for this paper by its authors.
  Use permitted under Creative Commons License Attribution 4.0
  International (CC BY 4.0).}

\conference{Fifth Knowledge-aware and Conversational Recommender Systems (KaRS) Workshop @ RecSys 2023, September 18--22 2023, Singapore.}

\title{VideolandGPT: A User Study on a Conversational Recommender System}


\author[1]{Mateo Gutierrez Granada}[%
orcid=,
email=Mateo.Gutierrez.Granada@rtl.nl,]
\cormark[1]
\fnmark[1]

\author[1,2]{Dina Zilbershtein}[%
orcid=,
email=zilbershtein.dina@maastrichtuniversity.nl]
\cormark[1]
\fnmark[1]
\address[1]{RTL Nederland B.V., Hilversum, The Netherlands}
\address[2]{Maastricht University, Maastricht, The Netherlands}

\author[1]{Daan Odijk}[%
orcid=,
email=Daan.Odijk@rtl.nl,
]

\author[2]{Francesco Barile}[%
orcid=,
email=f.barile@maastrichtuniversity.nl]

\cortext[1]{Corresponding author.}
\fntext[1]{These authors contributed equally.}

\begin{abstract}
This paper investigates how large language models (LLMs) can enhance recommender systems, with a specific focus on Conversational Recommender Systems that leverage user preferences and personalised candidate selections from existing ranking models. We introduce VideolandGPT, a recommender system for a Video-on-Demand (VOD) platform, Videoland, which uses ChatGPT to select from a predetermined set of contents, considering the additional context indicated by users' interactions with a chat interface. We evaluate ranking metrics, user experience, and fairness of recommendations, comparing a personalised and a non-personalised version of the system, in a between-subject user study. Our results indicate that the personalised version outperforms the non-personalised in terms of accuracy and general user satisfaction, while both versions increase the visibility of items which are not in the top of the recommendation lists. However, both versions present inconsistent behavior in terms of fairness, as the system may generate recommendations which are not available on Videoland.
\end{abstract}

\begin{keywords}
  ChatGPT \sep
  Conversational Recommender Systems \sep
  Video Recommendations \sep
  Fairness
\end{keywords}

\maketitle

\section{Introduction}
\vspace{-.3\baselineskip}


Recommender systems have revolutionized various industries such as e-commerce, media, and online advertising by providing customized experiences based on users' profiles and behaviors. Initially, content filtering was used to match users based on their preferred categories \cite{Naumov19}, but the development of collaborative filtering techniques such as matrix factorization (MF) has enabled more effective personalization \cite{Koren09, hu2008collaborative}. More recently, the development of attention mechanisms that efficiently connect encoder and decoder via Transformer blocks \cite{vaswani2017attention} represented a significant advancement in neural architectures, initially for natural language processing.
The emergence of Large Language Models (LLMs), such as BERT \cite{Devlin18}, and subsequently GPT-3 and chatGPT \cite{Radford2018ImprovingLU, radford2019language, brown2020language}, is a direct result of this breakthrough.



As the Transformer architecture gained popularity in other domains, recommender system scholars also saw potential in the attention mechanism \cite{donkers2017sequential, kang18}, recognizing the utility of sequential information \cite{sun2019bert4rec, chen2019behavior, Gutierrez21Videoland}. Breakthroughs in NLP research continued with the addition of new LLMs such PaLM \cite{Chowdhery2022PaLMSL} and LLaMA \cite{Touvron23}. These advancements have not gone unnoticed by researchers from diverse domains, including Recommender Systems which tried to incorporate LLMs in their toolbox \cite{Li2023GPT4RecAG, Liu2023IsCA}. Overall, LLMs hold great promise for improving the performance and capabilities of recommender systems.


LLMs bring several benefits to recommendation systems, including extensive knowledge, reasoning, natural language processing, and explainability, boosting user engagement and trust. They incorporate context, user preferences, and feedback, and transfer knowledge between domains, making them potent for creating accurate, explainable recommendation systems \cite{Cui2022M6RecGP}.

However, generating recommendations using LLMs is challenging for quite a few reasons \cite{Liu2023IsCA}: they are prone to generate incomplete, hallucinatory and biased results \cite{RAY2023121}, along with factually accurate but contextually inconsistent outcomes. Updating the parametric knowledge base and accommodating input token length are also significant challenges. Consequently, modern research often sees LLMs as summarization and reasoning engines rather than knowledge-based solutions for recommender systems, despite efforts to merge these approaches \cite{Pan2023UnifyingLL}.

This paper examines the impact of LLMs on a recommender system that can converse and reason within the users' context, using their preferences and a set of personalised candidates. The study involves users from RTL's Videoland\footnote{\url{https://www.videoland.com/}}, the largest Dutch video-on-demand (VOD) platform. The aim is to investigate through a user study the user experience of personalised recommendations in a conversational context, including situations where users explicitly state their preferences using natural language. We examine whether there is a discernible difference in users' personalised and non-personalised LLM recommendations. Furthermore, we aim to determine if users are exposed to titles beyond the top ranking.

In addition to its focus on recommendation accuracy and performance, this study evaluates the safety and fairness of recommendations generated by our proposed Conversational Recommender System (CRS). We analyze if the LLM adheres to fairness definitions proposed by the research community \cite{Fairness1}. Adopting the principle of fairness as "no harm", it becomes evident that recommending items not accessible on the Videoland platform undermines the platform's interests by encouraging people to find relevant content somewhere else. In this context, our analysis prioritizes aligning our recommender system with Videoland's objectives and avoiding any adverse impact on the platform's operations and goals.

\label{sec:introduction}



\section{VideolandGPT}
\vspace{-.3\baselineskip}

\begin{figure}[t] 
\centering
\vspace{-.3\baselineskip}
\includegraphics[width=\columnwidth]{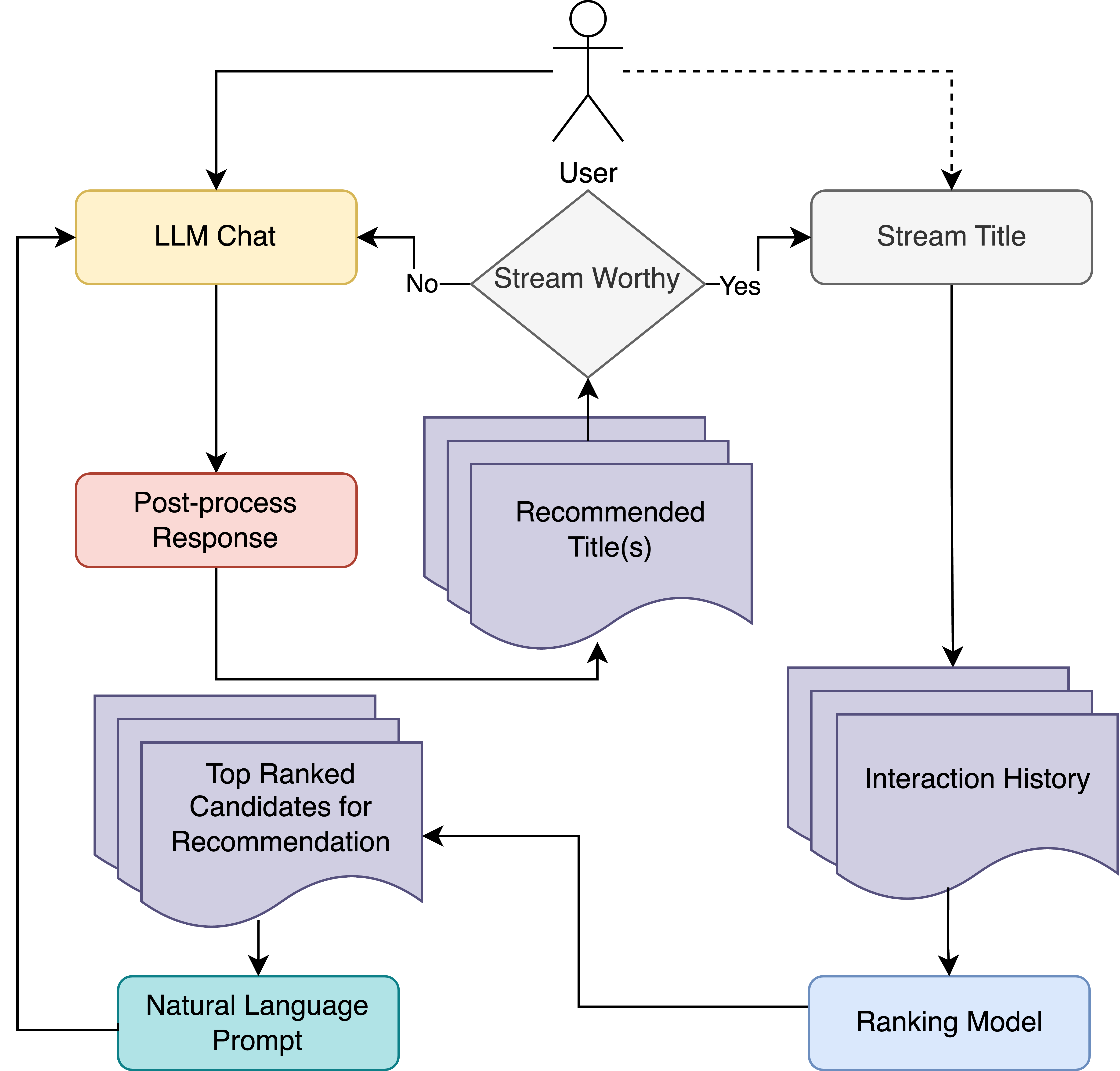}
\caption{VideolandGPT offers a direct feedback loop with the user, starting with their streaming activity and generating an Interaction History that serves as input to the Ranking Model. The primary filtering is in the Ranking Model, where the top $k$ titles are ranked and embedded as Candidates for Recommendation in a Natural Language Prompt. The LLM Chat is instructed to recommend titles from this subset during its conversation with the user. The post-processed response generated by the LLM Chat constitutes the Recommended Title(s), which the user can accept as a successful recommendation and subsequently stream, or continue the conversation with the LLM Chat.}
\label{fig:vlgpt}
\end{figure}


We evaluate our approach on a prototype conversational recommender system for Videoland, that we detail in this section. We base our prototype on the Ranking Model that we presented in \cite{Gutierrez21Videoland}. The architecture used to integrate the Ranking Model with the LLM's knowledge and capabilities, is illustrated in Figure \ref{fig:vlgpt}.   


In this architecture, the Ranking Model is considered a critical component of the solution and also a modular building block that can be replaced as needed. In our case, the model is an ensemble comprising a matrix factorization component \cite{hu2008collaborative} and a neural component \cite{Gutierrez21Videoland}, which utilizes the attention mechanism and sequential information in the \emph{Interaction History}. Our Ranking Model retrieves the top 300 titles for each user, reducing the catalog by approximately 90\%. We believe this number achieves a balance between relevance and discoverability.


\begin{figure}[t] 
\centering
\includegraphics[width=\linewidth]{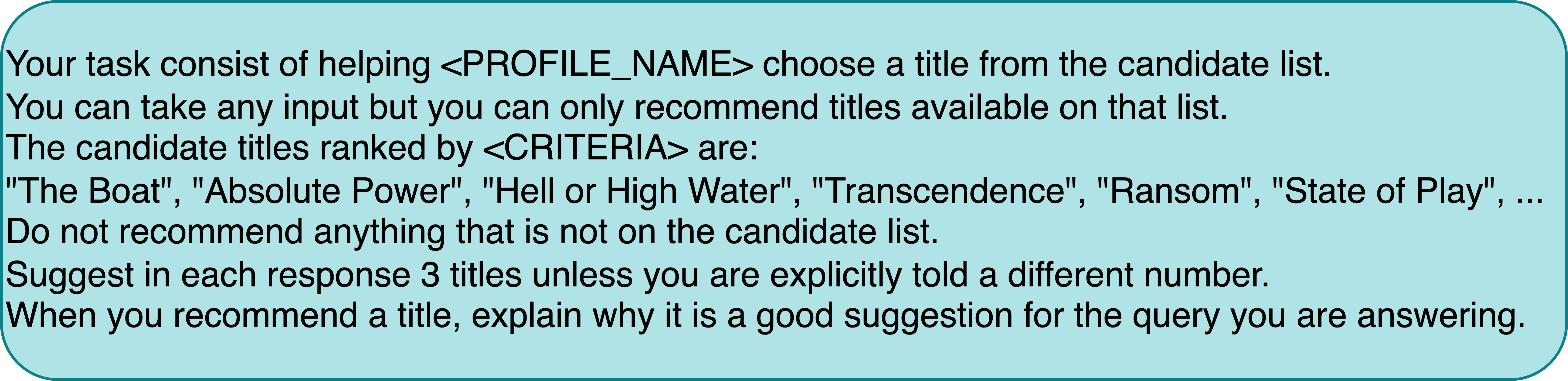}
\caption{This figure provides an example of a personalised prompt that tasks the LLM model with recommending three items to a specific user from a candidate list. The candidate list is personalised and can be dynamically sorted based on varying criteria.}
\label{fig:prompt}
\end{figure}

Our Natural Language Prompt is created to give precise instructions to the LLM Chat model to recommend titles from Videoland's candidates. We specify that the model should retrieve three items and provide explanations for each recommendation to improve explainability. An example of the prompt is illustrated in Figure \ref{fig:prompt}, which takes a candidate list of items sorted based on particular criteria and the user profile for which the recommendations are intended. This approach enables flexibility in accommodating various ranking methods.


The LLM Chat is designed to suggest a list of items that best matches the user's query and the candidate list of recommendations. As the conversation progresses, the user can either accept a recommendation or give feedback to the system to refine their discovery preferences. The user can also request new titles, ask for explanations for a particular recommendation, or seek further information related to it. We are testing our prototype with gpt-35-turbo as the LLM.


The post-processing step serves two critical functions. First, it enriches the LLM Chat's response with relevant metadata, such as the title's artwork and a direct link to stream it on Videoland. This additional information enhances the user's experience and makes it easier to access and enjoy recommended titles. Second, the post-processing step acts as a safeguard to remove any recommended title that is not directly aligned with the candidates for recommendation. In our experiment, we intentionally omitted this safeguard to examine the potential impact on platform fairness of not removing any recommended title that is offered by other platforms. 

\label{sec:method}

\section{User Study}
\vspace{-.3\baselineskip}
We conduct a small-scale user study to evaluate the performance of the recommender system. We compare two versions of the recommender system: a personalised version, based on users’ recommendations and a non-personalised one, based on the most popular titles. The study aims to assess user satisfaction, platform fairness aspects and to answer the main research questions: How can LLMs enhance (our) recommender systems? 
Can such a system, converse and reason within the user's context, using their preferences and a set of personalised candidates? 
Is a personalised chat-based recommender system perceived to be more enjoyable and more relevant compared to its non-personalised counterpart?

In a separate study, Radensky et al. \cite{Radensky23} examined the impact of confidence signal patterns on user trust and reliance in a music CRS. Their research inspired our evaluation approach, although our study covers broader aspects beyond confidence signals.

\paragraph{Participants}

The assignment of random groups was done prior to the study. The participants comprised employees within RTL. In total, 27 out of 42 invited participants took part in the study, ages ranging from 26 to 48, being 35\% of them women. Participation in the survey was voluntary, and the employees had not previously interacted with VideolandGPT. The sole requirement for participation was that the respondents must have watched at least one title on Videoland within the last 6 months to have personalised recommendations.

\paragraph{Experiment Protocol}

\begin{figure}
    \centering
    \includegraphics[width=\columnwidth]{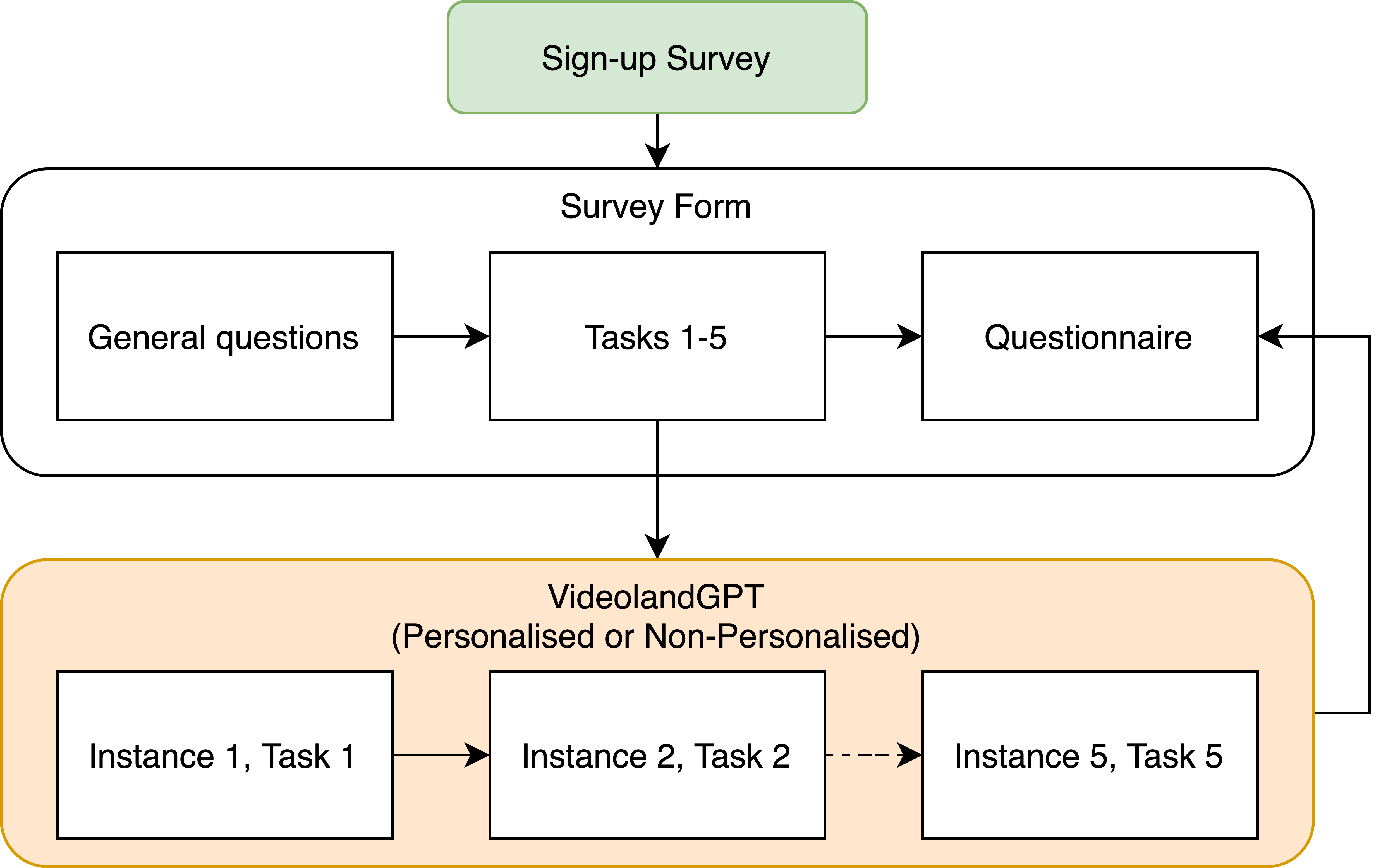}
\caption{Experiment scheme. Each participant signed up for the survey and subsequently completed the form, which included task descriptions and the questionnaire.}
\label{fig:experiment}
\end{figure}

The experiment's design is presented in Figure \ref{fig:experiment}. All study participants were explicitly requested to engage with the system in English throughout the study. Following this, the respondents were randomly divided into two groups. Participants, unaware of the version they were using, engaged with either a personalized or non-personalized VideolandGPT, the latter featuring top popular titles from Videoland's collection, ensuring unbiased results.


\begin{figure*}[t]
\centering
  \resizebox{1\textwidth}{!}{\includegraphics[width=0.8\linewidth]{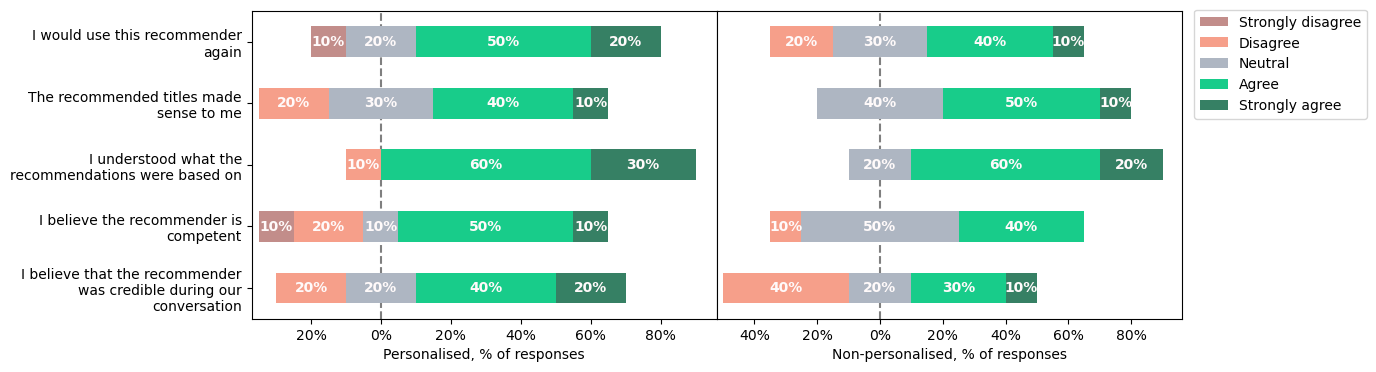}}
\vspace*{-\baselineskip}
\caption{Likert-type questions \cite{Toader2019TheEO}  with the results for both versions of the Recommender System, displaying the users' responses to the evaluation questions. The Likert scale was used to assess users' perceptions and satisfaction.}
\label{fig:questions}
\end{figure*}

Each participant was assigned a set of five tasks with the specific structures provided for each of them to ensure a more standardized evaluation process. Descriptions of the tasks are provided in Table \ref{tab:tasks}. However, participants were informed they could use their own words during interactions with the system, promoting natural conversation. The study was conducted online over a designated four-day period, offering convenience and flexibility to participants.

\begin{table}
\caption{Task description with the suggested requests for recommendations.}
\label{tab:tasks}
\begin{tabular}{p{0.15\linewidth} | p{0.73\linewidth}}
\toprule 
Task & Suggested initial prompt\\
\midrule
Title & Show me the most relevant titles considering that I like <TITLE>. \\
Topic & Show me the most relevant titles based on my passion for <TOPIC>.\\
Emotion & Show me the most relevant titles that will make me feel <EMOTION/DESIRE>.\\
Context & Show me the most relevant titles to watch with <GF/BF/SON/FRIEND> on a <DAY OF THE WEEK and/or EVENING/AFTERNOON/MORNING>.\\
Free & <Ask for 3 items to be recommended in any form you would like.>\\
\bottomrule
\end{tabular}
\end{table}


Assessment of the system was based on diverse forms of describing users' preferences, which included previously loved titles, topics, current or desired emotions, preferred company for movie-watching, and free-form requests. During the conversations, users had the opportunity to request the system to refine the recommendations twice, resulting in a maximum exposure to 9 items per task. Each task was completed in separate instances of the same version of the recommender, ensuring an isolated examination.

At the end of each task, respondents specified the title they considered the most relevant recommendation for them or stated that they did not receive a satisfactory recommendation. This feedback was used to understand VideolandGPT's recommendation capabilities, accuracy and fairness to the platform of the recommender.

Furthermore, because the participants were not exposed to VideolandGPT previously and to ensure the experiment's integrity, the order of the tasks was changed every five collected responses. By varying the task order, we sought to avoid any systematic influence on participants' responses, ensuring that the respondents' reactions to the tasks remained impartial and unaffected by the sequence in which they were presented.

After completing the tasks, participants were directed to fill out the questionnaire. The results to the Likert questions are  presented in Figure \ref{fig:questions}. Moreover, participants were asked to rank the tasks based on their satisfaction from the conversation with the recommender and were encouraged to provide any additional feedback they had regarding its use.
In addition, participants were asked about their native language, to explore any potential correlation between the quality of recommendations and their language background. This question was particularly relevant, as Videoland's collection primarily consists of contents in Dutch (57\% of the titles accounting for 63\% of the total available minutes).


\section{Evaluation}
\vspace{-.3\baselineskip}
We evaluate this study both quantitatively and qualitatively by analyzing the data collected from the logs of the conversation and the received questionnaire answers. It is important to note that not all responses from the conversations yielded usable data due to various reasons such as incomplete or ambiguous queries. As a result, we obtained 50 valid observations for each version of the recommender (five per respondent, one for each task).

\paragraph{Difference between two versions of the conversational recommender}

\begin{table*}
\caption{Experimental results on the two Ranking Models. The best results overall and per task are in boldface.
}
\label{tab:results}
\footnotesize
\resizebox{\textwidth}{!}{%
{\fontsize{7.4}{7.4} %
  \centering
  \begin{tabular}{llcccccc}
\toprule

\multirow{2}{*}{\bf Task} & \multirow{2}{*}{\bf Ranking Model} &  \multirow{2}{*}{\bf nDCG@9} & \multirow{2}{*}{\bf HR@9} & \bf Recommended & \bf Chosen in & \bf Chosen but not & \bf Unique Titles \\
    &  &  &  & \bf  in Candidates & \bf Candidates & \bf in Candidates & \bf per User \\
    \hline
    \midrule
    \multirow{2}{*}{\bf Overall} 
    & \bf Personalised & \bf0.4273 & \bf0.78 & 0.6958 & \bf0.54 & 0.24 & \bf24.9 \\
    & Non-personalised & 0.3880 & 0.74 & \bf0.7636 & 0.52 & \bf0.22 & 26.2 \\

    \bottomrule
    \multirow{2}{*}{\bf Title} 
    & Personalised & 0.3537 & 0.60 & 0.5750 & 0.40 & 0.20 & 5.6 \\
    & Non-personalised & \bf0.5635 & 0.80 & 0.7188 & 0.70 & 0.10 & 5.1  \\
    
    \cline{2-8}
    \multirow{2}{*}{\bf Topic} 
    & Personalised & 0.4848 & 0.80 & 0.4833 & 0.30 & 0.50 & \bf4.3 \\
    & Non-personalised & 0.4438 & 0.70 & 0.6722 & 0.30 & 0.40 & 5.4 \\

    \cline{2-8}
    \multirow{2}{*}{\bf Emotion} 
    & Personalised & 0.3421 & 0.70 & 0.8355 & 0.60 & 0.10 & 6.6 \\
    & Non-personalised & 0.2185 & 0.60 & \bf0.9380 & 0.60 & \bf0.00 & 7.3 \\

    \cline{2-8}
    \multirow{2}{*}{\bf Context} 
    & Personalised & 0.4215 & \bf0.90 & 0.9222 & \bf0.80 & 0.10 & 5.9 \\
    & Non-personalised & 0.4371 & \bf0.90 & 0.8611 & 0.60 & 0.30 & 5.2 \\

    \cline{2-8}
    \multirow{2}{*}{\bf Free} 
    & Personalised & 0.5343 & \bf0.90 & 0.6633 & 0.60 & 0.30 & 4.6 \\
    & Non-personalised & 0.2772 & 0.70 & 0.6277 & 0.40 & 0.30 & 6.7 \\

    
\bottomrule
\end{tabular}
}%
}
\end{table*}

Table \ref{tab:results} presents the metrics used to evaluate both versions. We measured  accuracy and relevance of recommendations by allowing participants to interact with 3, 6, or 9 recommended titles (with 8\% of sessions interacting with other numbers < 9) in our experiment. We evaluated the recommendations' performance using nDCG@9 and HR@9 metrics, considering all participants regardless of the number of titles they interacted with.

The personalised framework demonstrated a 10\% relative improvement over the non-personalised version in all tasks, highlighting the effectiveness of chat-based recommendations in improving user satisfaction and relevance in our research context.

To assess the fairness of the recommender system to the platform and its compliance with the rules, we measured the proportion of recommended and chosen titles that were in the candidate list and the chosen titles that were not on the candidate list. Moreover, we consider a measure of efficiency the number of unique titles recommended per user. While the personalised recommender outperformed in relevance metrics, our examination revealed inconsistencies in fairness metrics. For both recommenders, over 22\% of tasks had user-selected recommendations that were not available on Videoland, suggesting that the system occasionally generated recommendations beyond the platform's content availability, despite our attempts to control it.

\begin{figure}[t] 
\centering
\includegraphics[width=\columnwidth]{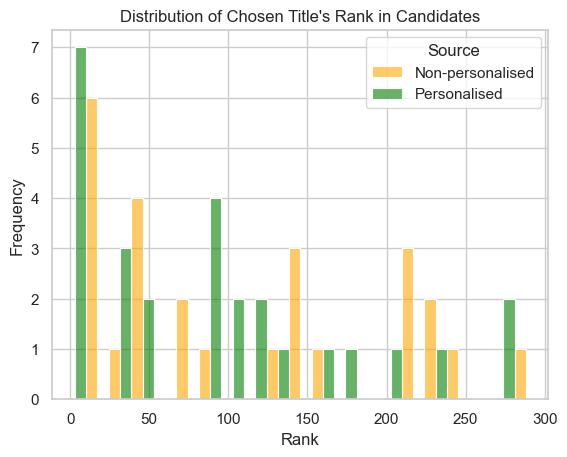}
\vspace{-1.5\baselineskip}
\caption{Distribution of the original rank in the candidates list for the titles that were chosen.}
\label{fig:dist_rank}
\end{figure}

Finally, the results presented in Figure \ref{fig:dist_rank} indicate how often users choose titles beyond the top ranking items. Our findings demonstrate that having a large pool of candidates is valuable, as users frequently select titles from across the entire range of recommendations.

\paragraph{Overall experience of using a conversational recommender}

In the second phase of our evaluation, we analyzed the feedback received from the questionnaire. The metrics substantiated the results, revealing a statistically significant positive correlation (Pearson coefficient of 0.26) between quantitative metrics like nDCG@9 and qualitative metrics like users' task rankings.
For instance, the \emph{Title} task was preferred by 30\% and 60\% of participants for personalised and non-personalised versions, respectively, in their rankings, aligning with corresponding nDCG scores. These findings endorse our metrics' effectiveness in capturing user preferences and judgments. However, it is important to note that this difference, while notable, is not statistically significant due to the relatively small sample size of participants. Consequently, providing an explanation for why the non-personalised version performed better on this task is challenging and requires further investigation.

The Likert questions answers indicate that a comparable proportion of respondents agreed or strongly agreed with three or more statements for both versions of the recommender system (70\% for personalised and 60\% for non-personalised). However, there is a notable difference: 40\% of respondents in the personalised version expressed agreement with all statements, while only 10\% did so in the non-personalised version. This suggests that the personalised version garnered a higher percentage of highly satisfied users with its recommendations. Additionally, we can observe that the non-personalised version elicited more neutral responses from the participants. This suggests a more mixed perception of the non-personalised version's recommendations.

The findings from the open-ended questions shed light on user perceptions of the recommender system's experience. Notably, 80\% of users perceived the personalised version as enjoyable, even when their specific requests were not entirely met. In contrast, 60\% of users found the non-personalised version enjoyable despite similar circumstances. The respondents also mentioned, that this recommender \emph{"could bring added value to the Videoland experience"}. A common feedback from participants who expressed dissatisfaction with their experience was the unavailability of relevant titles on the platform.

\section{Discussion and Conclusion}
\vspace{-.3\baselineskip}
Our study demonstrated that the personalised recommender outperformed the non-personalised version by delivering more relevant recommendations to users. However, it's important to recognise that both versions of the recommender still, in some cases, suggested titles that were not available on the platform, contrary to our initial expectations. This aspect highlights the need for further improvements and considerations in ensuring system consistency.
Despite this drawback, the study shed light on the potential of personalised chat-based recommendations to improve user satisfaction and relevance, offering valuable insights for future developments in recommender systems.


Limitations of the study include a primarily Dutch-speaking sample (65\% of all of the participants) due to the platform catering to a Dutch-speaking population, limited sample size, and the need to consider privacy and user preferences when implementing conversational recommender systems. Furthermore, if users explicitly share personal details with a conversational recommender system, it could impact their comfort in utilizing the system. Safeguards must be in place to ensure safety and prevent users from exploiting the system.

In conclusion, the study emphasizes the potential of personalised chat-based recommendations to enhance user experience, but further research is required to develop a safer mechanism for LLMs usage, ensuring adherence to rules and understanding potential unfair scenarios.


    

\newpage
\small
\bibliography{References}

\end{document}